# Microscopic analysis of Li nuclei cluster photodisintegration reaction characteristics in broad energy range: from threshold until 100 MeV

**M A Zhusupov, K A Zhaksybekova and R S Kabatayeva**
Al-Farabi Kazakh National University, Almaty, Kazakhstan

E-mail: raushan.kabatayeva@gmail.com

**Abstract**
On the base of potential theory of light nuclei cluster photodisintegration the characteristics of Lithium nuclei cluster photodisintegration reactions are considered in the range of low and intermediate energies. At low energies the important role of E1-multipole and its interference with E2-multipole were considered. The essential point is the different character of interference of E1- and E2-amplitude for the direct and inverse reactions. If for the direct reaction the interference at scattering in forward semisphere has a constructive character, then in backward semisphere the interference of E1- and E2-amplitudes – is deconstructive one. For the inverse reaction the interference has an opposite character: in forward semisphere it is deconstructive, and in backward semisphere it is constructive. In the energy range above several MeV the E2-multipole becomes dominating. The effect of appearance of the node structure of the wave function of relative motion has been considered in the characteristics of Li cluster photodisintegration reaction with polarized and non-polarized photons.

## 1. Introduction

For the theory of photonuclear reactions the processes of two-particle photodisintegration of light self-conjugate nuclei (N = Z) with formation of particles with zero isotopic spin like $^4$He($\gamma$,d)d, $^6$Li($\gamma$,d)$\alpha$, $^{16}$O($\gamma$, $\alpha$)$^{12}$C and other are of peculiar interest. Cross sections of the reactions mentioned are extraordinarily small since according to selection rule by isotopic spin, the E1-transitions in the case of $\Delta T = 0$ are strongly suppressed and the E2-multipoles begin to play the determinative role, here a magnitude of the E2-multipole, in its turn, is defined by the kinematic factor of suppression which is included in operators of electromagnetic transitions $T^{el}_{J\lambda} \sim (k_\gamma r)^J$. By the reason of small values of cross sections the processes were investigated insufficiently from experimental point, particularly, in the near-barrier range.

For nuclear astrophysics the $\alpha\,d \to {}^6\text{Li}\,\gamma$ reaction represents a special interest as a unique source of formation of $^6$Li nuclei in the Big Bang [1]. Its study is important for thermonuclear applications as well. The matter is that a resonance in $\alpha$d-scattering at $\alpha$-particles energy in laboratory system of 2.109 MeV ($E_\alpha = 0.7$ MeV in system of inertia center) is an only process as a result of which $\alpha$-particles with energy less than 3.7 MeV (products of dt- and d$^3$He-synthesis) will effectively interact with the main components of dt- and d$^3$He-plasma in early generation thermonuclear facilities [2]. The resonance corresponds to the known level of $3^+$ in $^6$Li nucleus, a decay of which is accompanied with a typical radiation with $E_\gamma = 2.186$ MeV [3]. The cross section of $\alpha\,d \to {}^6\text{Li}\,\gamma$ reaction in this resonance equals 150 nb and it is one of the series of resonances suggested for γ-diagnostics of thermonuclear deuterium-tritium plasma [2].

As it is known in stellar nuclear processes the fundamental role is for quantum tunnel effect. With its help one can explain the radiation and evolution of stars and also the synthesis from the primary hydrogen of all elements of the periodic system. Particles in stellar interior participating in nuclear synthesis reactions have energies much less than the height of Coulomb barrier $B_C$. For instance, in the center of the Sun the average kinetic energy of particles is about 1 – 3 keV, whereas the Coulomb barrier for two protons is $B_C \approx 1$ MeV, at that the height of the barrier increases proportionally to the product of charges of particles participating in the reaction. In such a case only a tiny part of particles, which is the "tail" of the Maxwell's distribution participates really in nuclear reactions owing to the tunnel effect of penetrating through the potential barrier. It is turned out to be enough for stars to radiate the huge amount of energy during the billions of years!

Thus in nuclear reactions investigations the range of very low energies is important for astrophysical applications.

## 2. E1-transitions at low energies

Though the E1-transitions in $\alpha d \to {}^6Li\,\gamma$ reaction are forbidden by isospin selection rule, the breaking of symmetry of γ-quanta angular distribution with respect to $\theta = 90^0$ angle, typical for the case of "pure" E2-transitions, is an evidence of noticeable interference of the multipoles at low energies [4]. That is why the natural question is what reasons lead to appearance of E1-multipole. Like any quantum mechanical selection rules, the isospin selection rule is approximate one but it is realized with high order of accuracy. If one considers the isospin to be a good quantum number, that is to restrict with wave functions with T = 0, then usually in long-wave approach the two small corrections to the operator of E1-transition are considered: it is a spin part of the interaction operator and a term concerned with account of retardation effect. For the $\alpha d \to {}^6Li\,\gamma$ reaction which is under consideration these two corrections were turned out to be negligibly small [5].

There is one more reason for appearing of the E1-multipole, concerned with the pronounced αd-structure of $^6$Li nucleus, in consequence of which for the subsystems the charge center does not coincide with the mass center of the system. Indeed, for nucleus with *A* nucleons, consisting of *a* and *b* subsystems, the dipole operator is $\vec{d} = \sum_i^z \hat{e}_i(\vec{r}_i - \vec{R})$, here $\vec{R}$ – coordinate of the total system mass center. The last expression can be represented in a form of sum $\vec{d} = \vec{d}_a + \vec{d}_b + \vec{d}_\rho$, where $\vec{d}_a$ and $\vec{d}_b$ – are dipole operators acting in each of the subsystems, and

$$\vec{d}_\rho = e\vec{\rho} \cdot m_a m_b / (m_a + m_b) \cdot (Z_a/m_a - Z_b/m_b) \qquad (1)$$

Here $\vec{\rho}$ – is a coordinate of relative motion of clusters *a* and *b*. Using this formula for calculation of $\alpha + d \to {}^6Li + \gamma$ reaction, one obtains that in αd-system because of the difference in mass $m_\alpha - 2m_d \neq 0$ there appears a small dipole moment $d_\rho = 4.3 \cdot 10^{-4} \cdot e\rho$, which leads to appearance of E1-transition both in the radiation capture of α-particles and in the photodisintegration process of $^6$Li nucleus by α-particle and a deuteron channel [6].

Thus, in [6] the authors suggested the following form of the operator of E1-transition constructed with account of the corrections mentioned above

$$\hat{T}_{1\lambda} = \frac{2}{3}\sqrt{\frac{\pi}{6}} iek_\gamma^2 \frac{\mu_p + \mu_n}{M_N} \rho \left[\vec{Y}_1 \otimes \vec{S}_1\right]_{1\lambda} - \frac{1}{45}\sqrt{\frac{4\pi}{3}} iek_\gamma^3 \rho^3 Y_{1\lambda}(\Omega_\rho) + \sqrt{\frac{4\pi}{3}} iek_\gamma \frac{m_\alpha - 2m_d}{m_\alpha + m_d} \rho Y_{1\lambda}(\Omega_\rho). \qquad (2)$$

Note that the total cross sections of photodisintegration and radiation capture are connected with each other by the principle of detailed balancing [7, 8].

When calculating the cross sections for direct and inverse ${}^6Li\,\gamma \rightleftarrows \alpha d$ reactions the $^6$Li nucleus ground state wave function was chosen in three-body αnp-model [9]. The weight of αd-component for the wave function composed $P_{\alpha d}$=0.71. According to Pauli principle requirement the relative motion wave function $R_{2S}$ has a node in the inner part of the nucleus. A contribution of the D-component, weight of which is small, is not taken into account. When constructing the wave function of αd -scattering the authors used a deep potential with forbidden states modified with account of the spin-orbital interaction in the following way

$$V_0 = V_{00} + \Delta V \cdot (-1)^l + V_1 \cdot (\vec{l} \cdot \vec{s})$$

$$V_{00} = 76.73\,\text{MeV};\ \Delta V = 2.5\,\text{MeV};\ V_1 = 3.305\,\text{MeV};\ R_0 = 1.85\,\text{fm};\ a = 0.71\,\text{fm};\ R_c = R_0. \tag{3}$$

In figure 1 there are results of calculation of αd-scattering phase shifts. It is obvious, that d-phases have a resonance structure corresponding to the low-lying $3^+$, $2^+$ and $1^+$ levels with isospin T = 0 in the $^6$Li nucleus, what defines the character of energy dependence of $^6\text{Li}\,\gamma \rightleftarrows \alpha\,d$ processes' cross sections as one can see in figure 2.

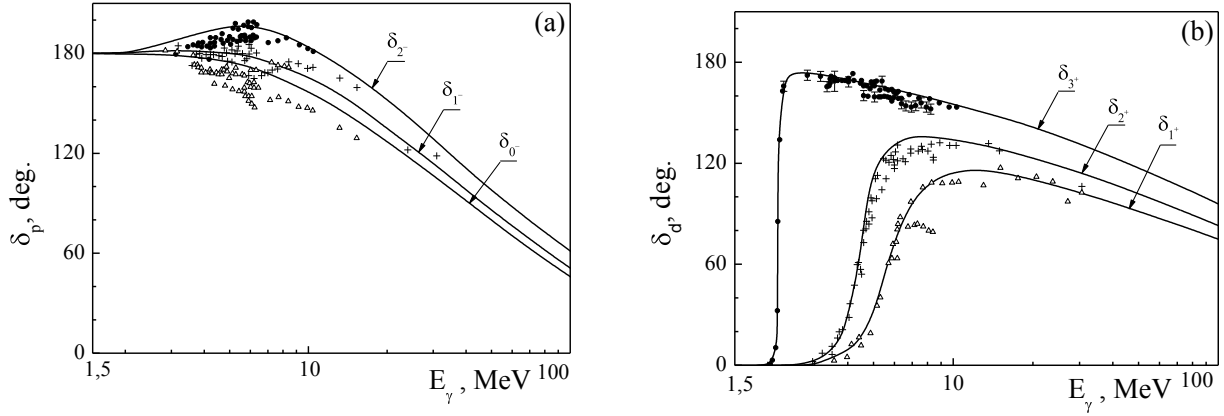

Figure 1 – Phase shifts of elastic αd-scattering.
Experimental data – from [10]. Theoretical calculation – with potential (3).

Note that the position of the first maximum in the cross section (figure 2) corresponds exactly to the narrow $3^+$-resonance at $E_\gamma = 2.186$ MeV. The second, wider maximum – is an "exit" from the range of $3^+$-resonance and a contribution of the soft overlapping $2^+$- and $1^+$-resonances. In the range of low energies $E_\gamma \sim 1.5$ MeV the contribution of E1-multipole is greater for an order of magnitude than the E2.

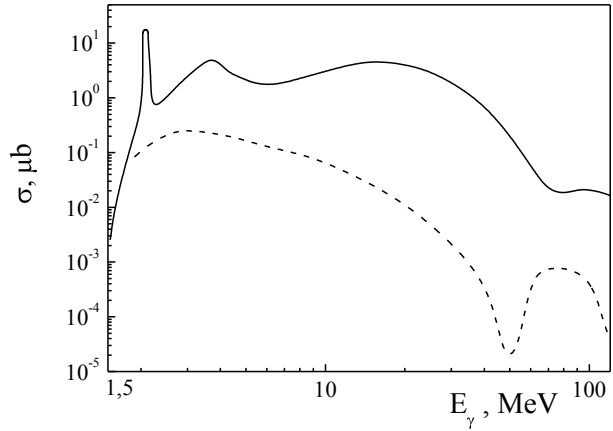

Figure 2 – Total cross section of $^6\text{Li}\,\gamma \to \alpha\,d$ reaction.
Dashed curve – pure E1-transition, solid curve – total result with account of E1- and E2-multipoles.

One can observe the E1-transition in angular distributions of $^6\text{Li}\,\gamma \rightleftarrows \alpha\,d$ processes in interference with E2-transition. In figure 3 there are our calculations carried out in αd-cluster model. As it is seen from figure 3b, the theoretical calculation is in qualitative agreement with experimental data on angular distributions [4].

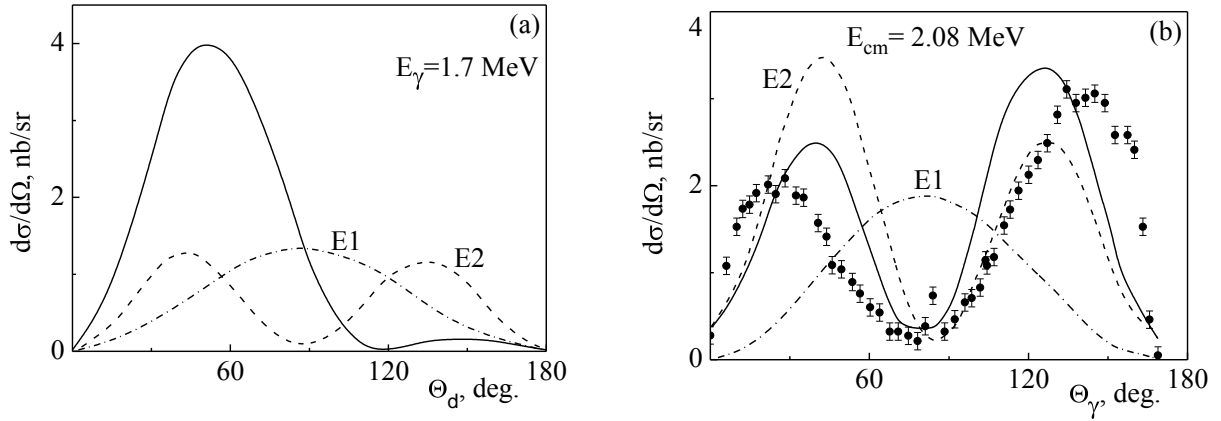

Figure 3 – Angular distributions of processes: (a) – $^6Li\,\gamma \to \alpha\,d$; (b) – $\alpha\,d \to {}^6Li\,\gamma$.
Dashed curve – pure E2-transition, dash-and-dot curve – E1-transition, solid curve – total result with account of E1- and E2-multipoles. Experimental data – from [4].

Focus an attention on the different character of E1- and E2-amplitudes' interference for direct $^6Li\,\gamma \to \alpha\,d$ and inverse $\alpha\,d \to {}^6Li\,\gamma$ reactions. If for the direct reaction the interference at scattering in forward semisphere (until $\pi/2$) have a constructive character, then in backward semisphere the interference of E1- and E2-amplitudes – is deconstructive. For the inverse reaction the interference has an opposite character: in forward semisphere it is deconstructive, but in backward semisphere it is constructive.

### 3. Angular distributions and asymmetry of deuterons

As an interesting characteristic for experimental investigations one can suggest calculations of asymmetry of angular distributions of deuterons in the $^6Li\,\vec{\gamma} \to \alpha\,d$ reaction with linearly polarized photons – see figure 4. Since the asymmetry $\Sigma$ is a relative characteristic then at good accuracy of experiment one can observe qualitative peculiarities in $\alpha d$-channel: resonance structure of E2-multipole transition and a mixture of E1-multipole as well.

Note that the appearance of the first minimum in $\Sigma$ is connected exactly with account of E1-multipole. At that our calculations show that such a peculiarity appears at angles $\theta_d \leq 110^0$. Thus the results obtained can serve as a good point for experimental set up with linearly polarized photons. A similar effect can be observed as well in experiments on measurement of polarization of secondary deuterons in the process of Coulomb dissociation of $^6Li$ nucleus in field of heavy nuclei.

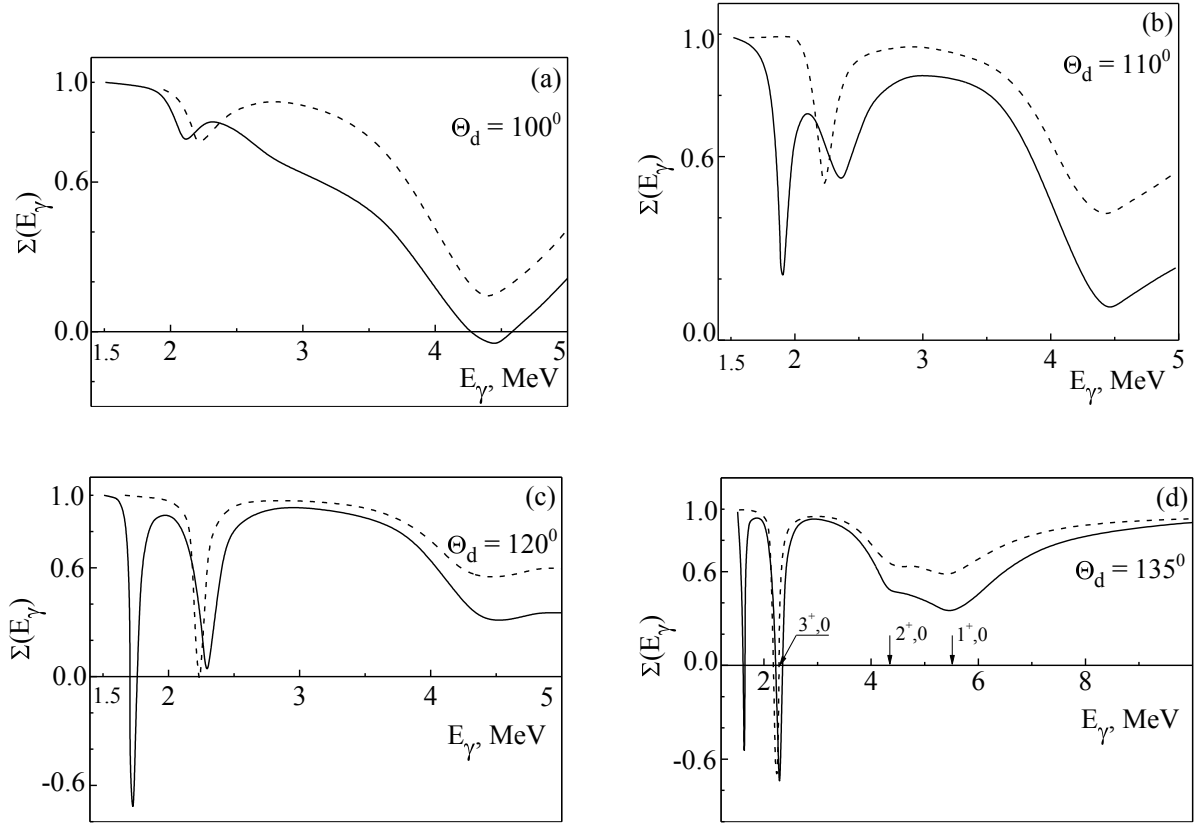

Figure 4 – Energy dependence of asymmetry in $^{6}\text{Li}\,\vec{\gamma} \to d\,\alpha$ process with linearly polarized photons at $100^0 < \theta_d < 135^0$. Theoretical calculation: dashed curve – pure E2-transition; solid curve – total result with account of E1- and E2-multipoles.

Thus, the main dominating transition in the $^{6}\text{Li}\,\gamma \rightleftarrows \alpha\,d$ processes is the quadrupole E2-transition, and the dipole isoscalar E1-transition only can be the dominant correction to it.

In figures 5 and 6a there are comparisons of the calculated angular distributions and the total $\sigma(E_\gamma)$ cross section with experimental data [11]. As it is seen from figure 5 in order to obtain a good agreement with an experiment the theoretical calculation should be multiplied by the factor 3. In our opinion in order to reproduce the available experimental data for the total cross section of the $^{6}\text{Li}(\gamma,d)\alpha$ process under consideration it is necessary to take into account the mixture of D-component of $^{6}\text{Li}$ nucleus wave function and a contribution of the matrix element of g-wave in the result of the quadrupole E2-transition $g \xrightarrow{E2} D$. It is worth to notice that the g-phase of elastic αd-scattering (figure 6b) in the range above 20 MeV – is smooth and one can suppose that the minimum in the total cross section at $E_\gamma \sim 70$ MeV (figure 6a) will be filled.

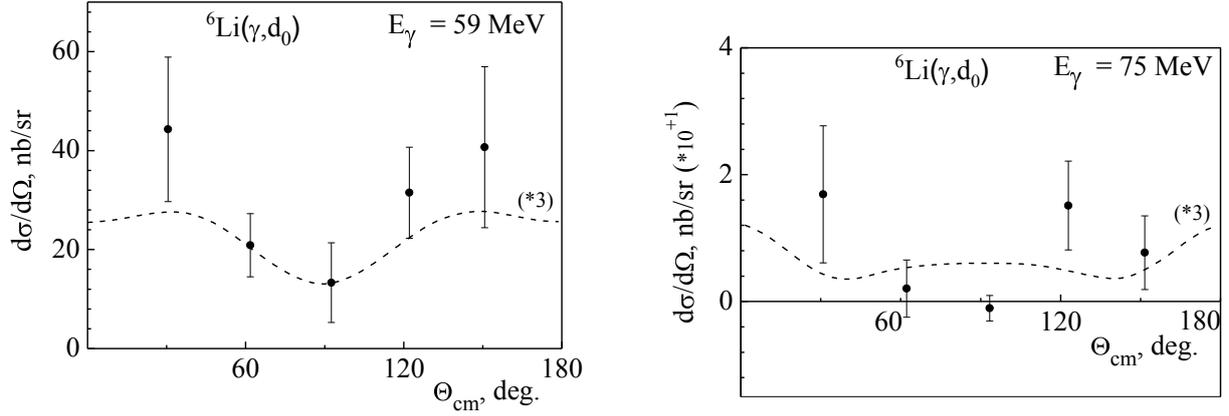

Figure 5 – Angular distributions of deuterons in $^6Li(\gamma,d_0)$ reaction at $E_\gamma$ = 59 and 75 MeV. Experimental data – from [11], dashed curve – theoretical calculation.

The cross section of the $^6Li\ \gamma \to \alpha\ d$ process under investigations are small by magnitude and that is why are difficult to measure. But the revealed effects of E1- and E2-multipoles' interference appear as well in the polarizing observables which are relative characteristics and that is why more accessible for measurement. As it is seen from figure 7, in the asymmetry $\Sigma(E_\gamma)$ there is observed a minimum in the range of $E_\gamma \sim 70$ MeV, origination of which has the same nature as the minimum in the total cross section, and exactly, node behaviour of a function of $\alpha$d-relative motion in $^6$Li nucleus.

Further, an analysis of figure 7 showed, that in forward semisphere $\theta_d = 60°$ there is no additional minimum appearing at $\theta_d = 135°$ because of the deconstructive interference of E1- and E2-multipoles.

Concerning the $\alpha\ d \to\ ^6Li\ \gamma$ reaction, the energy range below 700 keV is still important. In this range there are only data obtained in the result of Coulomb dissociation of lithium nuclei in field of heavy nucleus (Pb), that is using a virtual photon method [12].

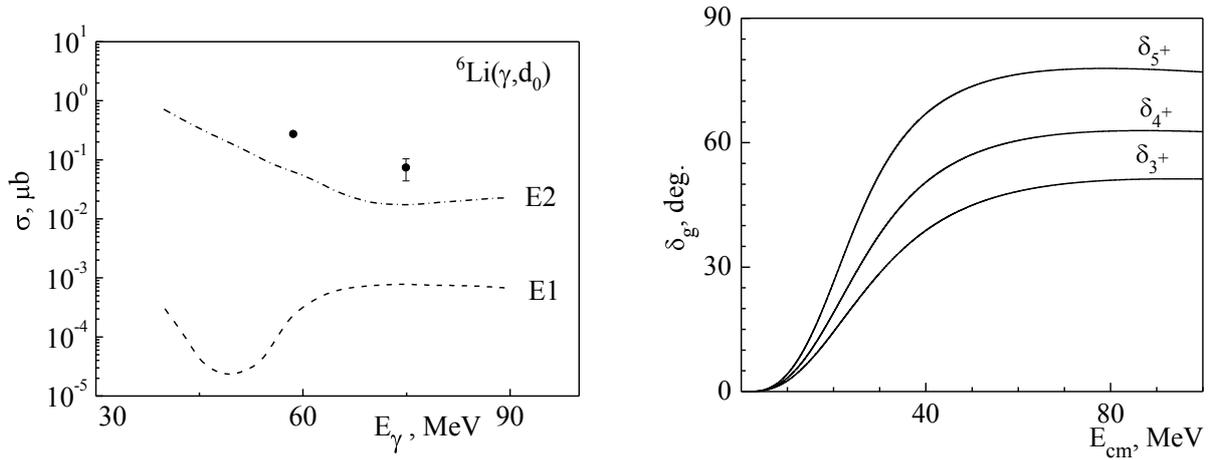

Figure 6 – (a) – total $\sigma(E_\gamma)$ cross section for $^6Li(\gamma,d_0)$ process. Experimental data – from [11], dashed curve – theoretical calculation; (b) – g-phase of elastic $\alpha$d-scattering.

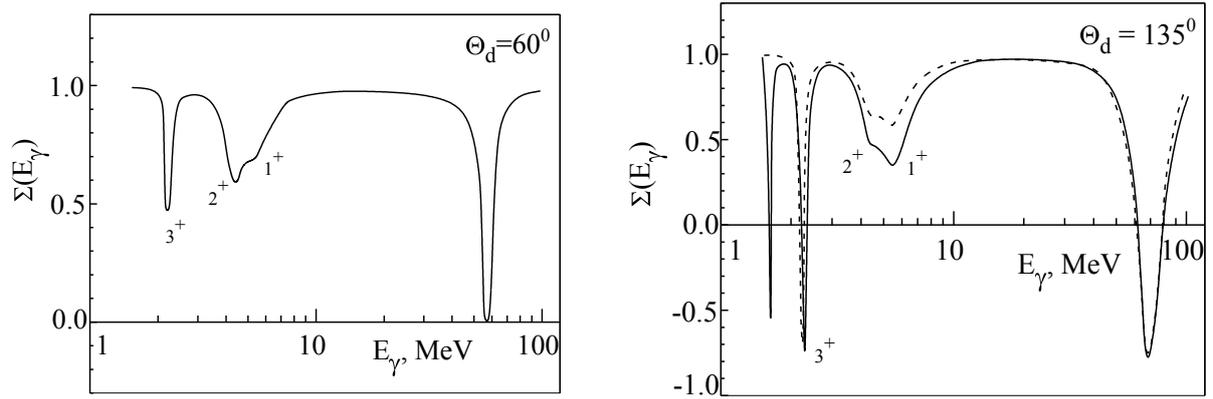

Figure 7 – Energy dependence of asymmetry in $^6$Li$(\gamma,d)\alpha$ process at $\theta_d = 60^0$ and $135^0$. Dashed curve – pure E2-transition, solid curve – total result with account of E1- and E2-transition.

The most interesting theoretical question – is a role of the E1-multipole. What its nature and magnitude are, and at what energies the dipole multipole will give a contribution comparable to the quadrupole one? To answer these questions it is necessary along with the angular distributions of deuterons to measure the total cross sections. Calculations carried out by physicists in 1991 [13] and other authors in 1995 [14] reproduce approximately equally the experimental data on total cross sections and astrophysical factor.

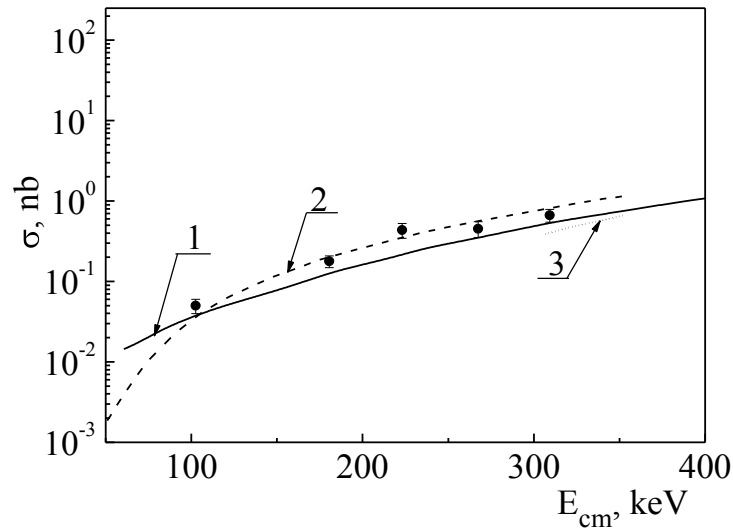

Figure 8 – Total cross sections of $\alpha\, d \to {}^6$Li $\gamma$ process. Experimental data – from [12].
Theoretical calculation: 1 – calculation from [13]; 2 – [14]; 3 – [15].

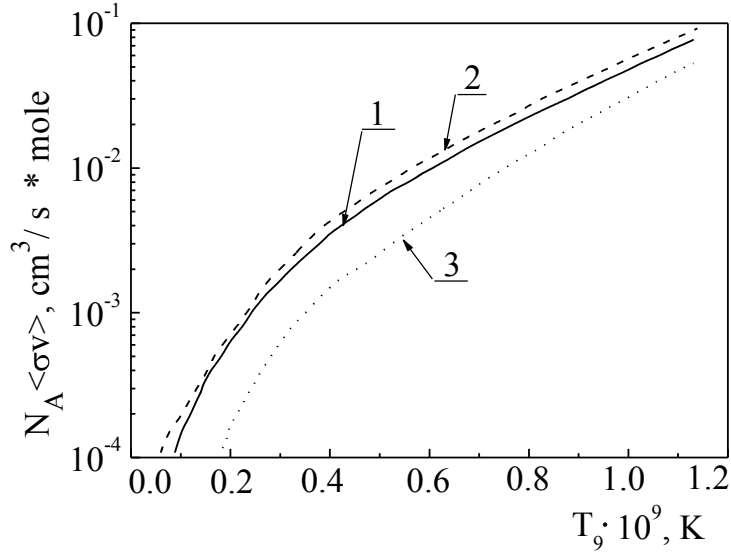

Figure 9 – Astrophysical $N_A \langle \sigma v \rangle$ velocities of $\alpha d \to {}^6Li\, \gamma$ reaction.
Theoretical calculation: 1 – calculation from [13]; 2 – [16]; 3 – [15].

In difference from the results of authors of work [14] the earlier results of other researchers [13] were theoretical predictions since they were carried out before their acquaintance with the experimental work [12].

The role of ddd-component in the wave function of $^6Li$ nucleus which was calculated in the MSU [17] is interesting. The weight of the component is small: on the level of 1%, but it can increase essentially the contribution of E2-multipole at very low energies at the expense of penetrability factor. By the same reason the significant M1-multipole can appear at low energies.

Note also that it is interesting to investigate the energy range of $E_\gamma \sim 40 \div 100$ MeV for the $^6Li + \gamma \to \alpha + d$ reaction both theoretically, with involvement of g-wave as a result of quadrupole E2-transition and account of D-component mixture in $^6Li$ nucleus wave function, and experimentally, what will allow getting new information about $^6Li$ nucleus clusterization.

**Conclusion**

The cluster E1-transition appears due to large difference in masses ($2m_d - m_\alpha$) that is large binding energy of α-particle in dd-channel equal to 24.5 MeV.

Because of the difference in potential barrier penetrability the E1-multipole appears in the astrophysical range, where the interference effects of E1- and E2-multipoles in the angular distributions of particles are the strongest ones.

At this the character of E1- and E2-multipoles' interference in the direct and inverse photonuclear reactions are different.

For the direct photonuclear reactions the analysis of angular distributions of particles has been enlarged with calculations of asymmetry appearing in case of processes with linearly polarized photons, exactly in this characteristic all peculiarities of (γ, d) reaction on $^6Li$ nucleus are revealed especially clearly: appearance of the cluster E1-multipole, its interference with E2-multipole; the node character of wave function of nucleus.